\documentclass[twocolumn]{autart}    % Enable this line and disable the
\usepackage{graphicx}          % Include this line if your
\usepackage{times} % assumes new font selection scheme installed
\usepackage{amsmath} % assumes amsmath package installed
\usepackage{amssymb}  % assumes amsmath package installed
\usepackage{subcaption}
\usepackage{setspace}

\usepackage{mathptmx} % assumes new font selection scheme installed
\usepackage{floatrow}
\usepackage{enumerate}
\usepackage{mathrsfs}
\usepackage{amsfonts}

\usepackage{xcolor}

\begin{document}

\begin{frontmatter}

\title{Quantum estimation, control and learning: opportunities and challenges\thanksref{footnoteinfo}} % Title, preferably not more

\thanks[footnoteinfo]{This work was supported by the Australian Research
Council's Discovery Projects funding scheme under Projects DP190101566, DP200101938 and DP180101805, and U.S.
Office of Naval Research Global under Grant N62909-19-1-2129.}
\author[UNSW-Canberra]{Daoyi Dong}\ead{daoyidong@gmail.com},
\author[ANU]{Ian R Petersen}\ead{i.r.petersen@gmail.com}
% (ead) as shown  % Add the           % e-mail address

%\thanks[c]{corresponding author.}

\address[UNSW-Canberra]{School of Engineering and Information Technology, University of New South Wales, Canberra, ACT 2600, Australia}
\address[ANU]{School of Engineering, The Australian National University, Canberra, ACT 2601, Australia}

                             % keyword list or with the
                                          % help of the Automatica
                                          % keyword wizard

\begin{abstract}                          % Abstract of not more than 200 words.
The development of estimation and control theories for quantum systems is a fundamental task for practical quantum technology. This vision article presents a brief introduction to challenging problems and potential opportunities in the emerging areas of quantum estimation, control and learning. The topics cover quantum state estimation, quantum parameter identification, quantum filtering, quantum open-loop control, quantum feedback control, machine learning for estimation and control of quantum systems, and quantum machine learning.
\end{abstract}

\end{frontmatter}

\section{Introduction}
The establishment of quantum mechanics is one of the greatest achievements in the 20th century. Quantum mechanics provides a framework to help us describe and understand the physical world involving microsystems such as electrons, photons and atoms, and is also crucial to help explain why stars shine and how the Universe formed. The unique characteristics of quantum systems have been found to be useful to provide advantages for developing powerful quantum technologies \cite{Dowling-and-Milburn-2003PTRSLA}. Emerging quantum technologies including quantum information technology, quantum sensing and quantum simulation are rapidly developing and various quantum platforms such as quantum optics \cite{Bachor-and-Ralph-2019book}, quantum superconducting systems \cite{Xiang-Ashhab-et-2013RMP}, spin systems \cite{Vandersypen-and-Chuang-2004RMP}, \cite{Khaneja-Brockett-et-2001PRA} have been used to develop quantum technology. Quantum simulation (i.e., simulating quantum mechanics) promises to have wide applications in e.g., condensed-matter physics, atomic physics, and molecular chemistry \cite{Buluta-and-Nori-2009Sci}. Quantum information technology (including quantum communication and quantum computation) has many important potential applications due to its advantages over traditional information technology. Quantum computation can take advantage of the unique quantum effects of entanglement and coherence to efficiently speed up the solutions of some classical problems and can even solve some difficult problems that classical (non-quantum) computers cannot \cite{Nielsen-and-Chuang-2000Book}, \cite{Shor-1994SFCC}. In quantum sensing, the principles of quantum mechanics can be effectively utilized to achieve highly sensitive sensing beyond the capabilities of classical sensors \cite{Degen-Reinhard-Cappellaro-2017RMP}. The ultimate limit on the accuracy of quantum sensing is set by quantum mechanics, and quantum resources (e.g., quantum entanglement) can be effectively exploited to enhance measurement precision \cite{Degen-Reinhard-Cappellaro-2017RMP}.

The great potential of quantum technology to promote economic and technological growth has been highlighted in reports and initiatives of various countries such as the National Quantum Initiative Act (US), ``The Quantum Age: technological opportunities" (UK) \cite{UK-roadmap}, European Quantum Technologies Roadmap \cite{Acin-Bloch-et-2018NJP}, Australia's Quantum Technology Roadmap \cite{Australia-roadmap} and ``Quantum information research in China" \cite{Zhang-Xu-Li-et-2019QST}. In the last several years, quantum computers have made great progress; see, e.g., Google Sycamore (53-qubit quantum processor) \cite{Arute-Arya-Babbush-et-2019Nat}, Jiuzhang (76-photon photonic quantum processor) \cite{Zhong-Wang-Deng-et-2020Sci}, IBM Eagle (127-qubit quantum processor) \cite{IBM-127}. The progress in quantum technology provides many exciting opportunities for systems and control researchers. For example, developing estimation and control theories for quantum systems is a fundamental task for practical quantum computation and quantum sensing. In this article, we briefly present some challenging open problems and potential opportunities in the emerging areas of quantum estimation, control and learning from the perspective of systems and control \cite{Dong-and-Petersen-2010IET}, \cite{Altafini-and-Ticozzi-2012TAC}, \cite{Glaser-Boscain-et-2015EPJD}, \cite{Brif-Chakrabarti-et-2010NJP}, \cite{Rabitz-Vivie-Riedle-et-2000Sci}.

This article is organized as follows. Section 2 introduces quantum state estimation, quantum parameter identification and quantum filtering. Section 3 focuses on quantum control including quantum open-loop control and quantum feedback control. Machine learning for quantum estimation and control, and quantum machine learning are discussed in Section 4. Concluding remarks are presented in Section 5.

\section{Estimation, identification and filtering in quantum systems}
\subsection{Quantum state estimation}
Quantum state estimation is often referred to as quantum state tomography in the quantum technology community. The basic aim of quantum state estimation is to reconstruct unknown quantum states. According to the quantum measurement postulate, the measurement on a quantum system usually modifies the state of this system, which makes the state estimation challenging. The state of a quantum system can be described by a density matrix $\rho$ satisfying $\rho^{\dagger}=\rho$, $\rho\geq 0$ and $\text{Tr}(\rho)=1$, where $\text{Tr}(\rho)$ returns the trace of $\rho$, and $\rho^\dagger$ is the conjugation ($*$) and transpose ($T$) of $\rho$. A pure state can also be described by a unit complex vector $|\psi\rangle$ with $\rho=|\psi\rangle\langle\psi|$ where $\langle\psi|=|\psi\rangle^{\dagger}$ \cite{Nielsen-and-Chuang-2000Book}. A mixed state is linear combination of independent pure states, satisfying $\text{Tr}(\rho^{2})< 1$. To estimate an unknown quantum state, we usually need to make measurements on many copies of the state. A quantum measurement can be described by a set $\{P_i\}$ of positive operator valued measurement (POVM) elements, where $P_i\geq 0$ and $\sum_{i}P_i=I$ with $I$ being the identity matrix. When making a measurement, the occurrence probability of the $i$th outcome can be calculated as $p_i=\text{Tr}(\rho P_i)$ according to the Born Rule. In quantum state estimation, we need to design a POVM measurement and develop an efficient estimation algorithm to reconstruct an unknown quantum state from measurement data. Various quantum state estimation methods have been developed such as the maximum likelihood estimation (MLE) method, the Bayesian mean estimation approach \cite{Paris-and Rehavek-2004Book} and linear regression estimation (LRE) \cite{Qi-Hou-Wang-et-2017npjQI}. In the MLE method, we choose the state estimate that gives the observed results with the highest probability. It is asymptotically optimal in the sense that the estimation error can asymptotically
achieve the Cram\'{e}r-Rao bound. The Bayesian mean estimation approach can always give a unique state
estimate, since it constructs a state from an integral averaging over all possible quantum states with proper
weights. In the LRE method, the reconstruction problem of a quantum state can first be converted into a parameter-estimation problem of a linear regression model and then the state can be estimated through solving the linear regression model \cite{Qi-Hou- Li-et-2013SR}. Among these three approaches, the most popular method is MLE while the most efficient approach is LRE for full reconstruction of quantum states. MLE usually involves solving a large number of nonlinear
equations where their solutions are usually difficult to obtain and often not unique. The Bayesian mean estimation method has high computational complexity which limits its application. An advantage of LRE is that we may characterize its computational complexity and obtain a theoretical error upper bound. Another advantage of LRE is that it is suitable for parallel computation \cite{Hou-Zhong-Tian-et-2016NPJ} and developing an adaptive quantum state estimation method \cite{Qi-Hou-Wang-et-2017npjQI}.

We use the LRE method as an illustrative example to explain the challenges of quantum state estimation \cite{Qi-Hou- Li-et-2013SR}, \cite{Dong-and-Wang-2017ANZCC}.
Consider a quantum system consisting of $n$-qubits associated with Hilbert space $\mathcal{H}$. Let $\{\Omega_{i}\}^{2^{2n}-1}_{i=1}$ denote a set of Hermitian operators satisfying (i) $\textmd{Tr}(\Omega_{i})=0$ and  (ii) $\textmd{Tr}(\Omega_i\Omega_j)=\delta_{ij}$, where $\delta_{ij}$ is the Kronecker function. The quantum state $\rho$ to be reconstructed can be parameterized as $$\rho=\frac{I}{2^{n}}+\sum^{2^{2n}-1}_{i=1}\theta_i\Omega_i,$$ where $\theta_i=\textmd{Tr}(\rho\Omega_i)$. Let $\Theta=(\theta_1, \cdots, \theta_{2^{2n}-1})^{T}$. Suppose a series of quantum measurements $\{P^{(j)}\}^{M}_{j=1}$ is performed. Then each operator $P^{(j)}$ can be parameterized under bases $\{\Omega_{i}\}^{2^{2n}-1}_{i=1}$ as \cite{Qi-Hou- Li-et-2013SR} $$P^{(j)}=\frac{\gamma^{(j)}_{0}}{2^{n}}+\sum^{2^{2n}-1}_{i=1}\gamma^{(j)}_{i}\Omega_i,$$
where $\gamma^{(j)}_{0}=\textmd{Tr}(P^{(j)})$ and $\gamma^{(j)}_{i}=\textmd{Tr}(P^{(j)}\Omega_i)$. Let $\Gamma^{(j)}=(\gamma^{(j)}_{1}, \cdots, \gamma^{(j)}_{2^{2n}-1})^{T}$. We can obtain the following linear regression equations for $j=1,\ \cdots,M$,
\begin{equation}\label{average2}
\hat{p}(P^{(j)})=\frac{\gamma^{(j)}_{0}}{2^{n}}+{\Gamma^{(j)}}^{T}\Theta+e^{(j)},
\end{equation}
where $\hat{p}(P^{(j)})$ is the estimate of the probability of obtaining the result of $P^{(j)}$, and $e^{(j)}$ denotes the estimate error which can be taken as noise. Once we obtain the solution $\Theta$, we may reconstruct the quantum state $\rho$. It is clear that both the dimension and the parameter number for the state to be reconstructed exponentially increase with the system size (i.e., the number of qubits). It is critical to develop efficient estimation algorithms for large-scale or many-body quantum systems. Another challenge is to optimize the estimation precision with limited resources (e.g., a given maximum number of copies, a limited set of measurements). Developing adaptive or learning methods may provide a solution to this problem. There are many other open problems in the area of quantum state estimation. For example, the efficiency of the estimation algorithms may be significantly enhanced when there is prior knowledge on the quantum state to be reconstructed. Different adaptivity criteria and the precision limit of adaptive estimation can be explored for adaptive estimation of quantum states. The capability of state-of-the-art machine learning for quantum state estimation is worth further exploring.

\subsection{Quantum parameter identification}
Identifying parameters is a fundamental task in science research and engineering applications. In parameter identification, the Cram\'{e}r-Rao bound is often used to characterize the precision limit. In quantum technology, parameter identification has close connection to and wide applications in quantum sensing and quantum control. As an illustration, assume that a single parameter to be identified $x$ is encoded in the dynamics of a quantum system $\Phi_{x}$. A probe state $\rho_{0}$ is prepared and it evolves to $\rho_{x}$ under the dynamics $\Phi_{x}$ which can be denoted as
\begin{equation}
\rho_{0}\ \ \ \underrightarrow{\Phi_{x}} \ \ \ \rho_{x}.
\end{equation}
The measurement is described by POVM $P$ over $N$ copies of the quantum system. The identification precision can be characterized based on the quantum Fisher information $\mathcal{F}(\rho_x)$ through the quantum Cram\'{e}r-Rao bound \cite{Yuan-2016PRL}
\begin{equation}
\Delta x \geq \frac{1}{\sqrt{N\mathcal{F}(\rho_x)}}
\end{equation}
where $\Delta x$ is the estimation uncertainty,
\begin{equation}
\mathcal{F}(\rho_x)=\max_{P}F(x)
\end{equation}
for classical Fisher information $F(x)$. One needs to optimize the Fisher information with respect to all possible POVMs to obtain the quantum Fisher information. The identification precision limit is related to the measurement scheme and the calculation of the quantum Fisher information is usually challenging. A fundamental precision limit in quantum parameter identification is Heisenberg limit and many studies have contributed to design estimation strategies to approach the precision limit (e.g., \cite{Hou-Jin-Chen-et-2021PRL}). Existing results show that entangled probe states and the introduction of proper Hamiltonian control may significantly enhance the identification precision \cite{Pang-and-Jordan-2017NC}. When multiple parameters to be identified, we need to use the quantum Fisher information matrix to characterize the quantum Cram\'{e}r-Rao bound and identification precision limit where non-commutative relationships between different parameters and measurements
pose new challenges and complexity (see \cite{Liu-Yuan-Lu-et-2020JPA} for a review of the quantum Fisher information matrix). Although a lot of work has been presented to investigate the identification problems of a single parameter and multi-parameters in quantum systems, there are still many open questions worth exploring, such as establishing the precision bound when various noises exist, how to balance the requirement of probe states and identification precision, how to design optimal control schemes to approach the precision limit for various systems, how to develop adaptive measurement and control strategies to enhance the identification precision, how to achieve high-precision parameter identification for open quantum systems, how to approach optimal estimate when the resources are limited, and how to implement high-precision parameter identification on various real quantum platforms.

Another fundamental task in quantum systems is to identify the system Hamiltonian $H$ since the Hamiltonian is an essential component to determine the dynamics of quantum systems. For example, the evolution of a closed quantum system can be described by
\begin{equation}
\dot{\rho}=-\frac{\text{i}}{\hbar}(\rho H-H\rho)
\end{equation}
where $\text{i}=\sqrt{-1}$, and $\hbar$ is the reduced Planck's constant. If we can identify the Hamiltonian $H$ and know the initial state $\rho_{0}$ of the system, the dynamics can be determined by $H$ and $\rho_{0}$. Various methods have been presented to investigate Hamiltonian identification problems.  For example, a Hamiltonian identification method using measurement time traces has been proposed based on transfer functions and classical system identification theory \cite{Zhang-and-Sarovar-2014PRL}. In \cite{Wang-Deng-Duan-2015NJP}, dynamical decoupling was employed for identifying Hamiltonian parameters and a two-step optimization Hamiltonian identification was presented in \cite{Wang-Qi-et-2018TAC}. Similar to the quantum state estimation task, a challenge in developing Hamiltonian identification algorithms is that the dimension and the parameter number for the Hamiltonian to be identified usually increase exponentially with the system size. This causes the resource requirement and the computational complexity of identification algorithms to quickly increase with the system size.

The most general quantum parameter identification problem in quantum technology is quantum process identification or quantum process tomography \cite{Nielsen-and-Chuang-2000Book}, \cite{Wang-Qi-et-2018TAC}. A quantum process $\mathcal{E}$ maps an input state $\rho_{in}$ to an output state $\rho_{out}$. In Kraus operator-sum representation \cite{Nielsen-and-Chuang-2000Book}, we have
\begin{equation}\label{kraus1}
\mathcal{E}(\rho_{in})=\rho_{out}=\sum_{i}A_i\rho_{in} A_i^\dagger,
\end{equation}
where $\{A_i\}$ is a set of $2^{n}\times 2^{n}$ matrices, with $\sum_i A_i^\dagger A_i\leq I$. Trace-preserving operations are usually considered, i.e.,
\begin{equation}\label{trpreserve}
\sum_i A_i^\dagger A_i= I.
\end{equation}
By expanding $\{A_i\}$ in a fixed family of basis matrices $\{E_i\}$, we obtain $A_i=\sum_j c_{ij}E_j$, and $\mathcal{E}(\rho_{in})=\sum_{jk}E_j\rho_{in} E_k^\dagger x_{jk}$, with $x_{jk}=\sum_i c_{ij}c_{ik}^*$. If we take matrix $X=[x_{ij}]=C^TC^*$ with $C=[c_{ij}]$. $X$ is called process matrix \cite{Wang-Qi-et-2018TAC} which has a one-to-one correspondence with $\mathcal{E}$. Hence, the full characterization of $\mathcal{E}$ can be completed by reconstructing $X$ and the completeness constraint (\ref{trpreserve}) corresponds to $\sum_{j,k}x_{jk}E_k^{\dagger}E_j=I$.
A central issue in quantum process tomography is to design an algorithm to find a physical estimate $\hat X$ close to the true $X$ by observing the output states for given probe states. Maximum likelihood estimation and Bayesian mean estimation can be used to achieve this task while their computational complexity is usually high for high-dimensional quantum systems \cite{Wang-Qi-et-2018TAC}. Designing more efficient algorithms is still an important open problem in quantum process identification.

Quantum system identification has rapidly developed in recent years \cite{Burgarth-and-Yuasa-2012PRL}, \cite{Wang-Yin-et-2019Auto}, \cite{Xue-Wu-Ma-et-2021PRA}. Many challenging problems are waiting for investigation. For example, although several results on identifiability of quantum systems \cite{Sone-and-Cappellaro-2017PRA}, \cite{Wang-Dong-et-2020TAC} have been presented, the identifiability of more general quantum systems is not investigated. The identifiability of quantum systems has a close connection to characterizing the capability of quantum sensors and may impose new challenges different from classical system identification problems \cite{Wang-Dong-et-2020TAC}, \cite{Yu-Wang-Dong-et-2021Auto}. Adaptive strategies have only been applied to several simple quantum identification tasks (e.g., estimating the Hamiltonian parameter of a two-level system \cite{Sergeevich-Chandran-Combes-et-2011PRA}) and more adaptive algorithms need to be developed to enhance the identification precision for quantum systems. Other new directions for future research include exploring distributed identification of quantum networks and enhancing the capability of quantum system identification using entanglement and squeezing.

\subsection{Quantum filtering and quantum smoothing}
When a quantum state or some parameters are dynamically changing, we usually employ the filtering theory to provide an estimate of the state or parameters. The theory of filtering often focuses on the estimation of the system states from noisy signals and/or partial observations. A filter is used to propagate knowledge about the system state given all observations up to the current time and provide an optimal estimate of the state.
Quantum filtering theory is extremely useful in developing measurement-based feedback control of quantum systems \cite{Wiseman-and-Milburn-2010Book}, \cite{Rouchon-and-Ralph-2015PRA}. An early approach to quantum filtering was presented by Belavkin \cite{Belavkin-1992JMA} who employed the framework of continuous nondemolition quantum measurement using the operational formalism. In the physics community, quantum filtering theory (called as quantum trajectory theory) was also independently developed in the early 1990s \cite{Carmichael-1993PRL}. Quantum filtering theory can be established based on quantum probability theory. In quantum probability theory, an isomorphic equivalence may be established between a commutative subalgebra of quantum operators on a Hilbert space and a classical (Kolmogorov) probability space through the spectral theorem. Any probabilistic quantum operation within the commutative subalgebra can be associated with its classical counterpart. For example, we consider an $n$-dimensional complex Hilbert space $\mathcal{H}$. Any self-adjoint operator $O$ on $\mathcal{H}$ has a spectral decomposition $O=\sum_{j=1}^n o_jP_{o_j}$ with the eigenvalues $\{o_j\}$ and orthogonal projection operators $\{P_{o_j}\}$ satisfying $P_{o_j}P_{o_k}=\delta_{jk}P_{o_k}$ and $\sum_{j=1}^n P_{o_j}=I$. For any continuous function $f: \mathbb{R}\to \mathbb{C}$, we have $f(O)=\sum_{j=1}^n f(o_j)P_{o_j}$. Thus the set $\mathscr{O}=\{X: X=f(O), f: \mathbb{R}\to \mathbb{C}\}$ forms a commutative $*-$algebra generated by $O$. $\mathbb{P}: \mathscr{O} \to \mathbb{C}$ denotes a mapping satisfying $\mathbb{P}(X)\geq 0$ if $X\geq 0$ and $\mathbb{P}(I)=1$. There is always a density operator $\rho$ such that $\mathbb{P}(X)=\text{Tr}(\rho X)$, where $\rho=\rho^{\dagger}, \text{Tr}(\rho)=1$ and $\rho \geq 0$. A commutative $*-$algebra structure is equivalent to a classical probability space. A pair $(\mathscr{N}, \mathbb{P})$ can be defined as a quantum probability space, where $\mathscr{N}$ is a $*-$algebra on $\mathcal{H}$ (see \cite{Bouten-van-Handel-James-2007SIAM} for more details).
Similar to the classical case, the optimal estimate of any observable is given by its quantum expectation conditioned on the history of continuous nondemolition quantum measurements. The quantum filter can be derived in terms of It\^{o} stochastic differential equations using a reference probability method (see, e.g., \cite{Bouten-van-Handel-James-2007SIAM}, \cite{Gao-Dong-Petersen-2016Auto} for more details). Quantum filtering theory has been widely applied in the analysis and control of quantum optical systems. More applications are worth exploring in various quantum systems, especially for the development of measurement-based quantum feedback control schemes. In these potential applications, the commutative $*-$algebra structure needs to be carefully designed to meet specific requirements since the possible non-commutation between different observables and/or between different time points for the same observable must be considered. Several challenging tasks include dimension reduction of quantum filters (especially for continuous variable quantum systems) \cite{Gao-Zhang-Petersen-2020Auto}, quantum filter design for quantum systems with both quantum random variables and classical random variables \cite{Tsang-2009PRL}, \cite{Gao-Dong-Petersen-2016Auto}, \cite{Yu-Dong-Petersen-et-22019TCST}, and quantum filtering for on-demand quantum feedback control.

In quantum filtering, the filter estimate at time $t$ is calculated based on the observation records before $t$. When real-time estimation is not required, we may develop quantum smoothing theory to further improve the estimation performance \cite{Chantasri-Guevara-Laverick-et-2021PR}. In quantum smoothing, the observation records both before (past records) and after time $t$ (future records) are used. Existing results have shown that the estimate based on quantum smoothing using both past and future records is statistically closer to the true values of the quantity under estimation than the filtered estimate \cite{Chantasri-Guevara-Laverick-et-2021PR}. The reason is that quantum smoothing uses more observation information than quantum filtering. As a quantum system's observables at $t$ may not commute with operators representing measurement results of the system at later time \cite{Chantasri-Guevara-Laverick-et-2021PR}, \cite{Wiseman-and-Milburn-2010Book}, a straightforward application of classical smoothing theory to quantum smoothing problems may lead to nonphysical estimated states. Hence, it is required to develop quantum smoothing theory to accommodate these unique properties different from classical smoothing problems. Several fundamental aspects of quantum smoothing theory have been investigated by considering various cost functions (see, e.g., \cite{Chantasri-Guevara-Laverick-et-2021PR}). Various engineering applications of quantum smoothing theory are worth exploring.

\section{Quantum control}
\subsection{Quantum open-loop control}
Feedback is one of the most fundamental concepts in classical systems and control, and closed-loop control is the most popular paradigm in classical control systems. However, the unique characteristics of quantum mechanics (e.g., measurement backaction) introduce new challenges for developing closed-loop and feedback control theory. Open-loop control is still a widely used paradigm in quantum control design.
In many practical quantum control applications, it is desired to design a control law to achieve a given objective with optimal performance. Optimal control theory and methods have been widely employed in
seeking an optimal control law \cite{D'Alessandro-2007Book},
\cite{Werschnik-and-Gross-2007JPB}.
In quantum optimal control, the control task is usually
formulated as a problem of searching for a set of admissible controls
to minimize a cost functional related to practical requirements. The cost functional may be the control time (time-optimal control) \cite{Khaneja-Brockett-et-2001PRA}, \cite{Sugny-Kontz-Jauslin-2007PRA}, the control energy
\cite{D'Alessandro-and-Dahleh-2001TAC}, the fidelity of the final state with a given target state, the sensitivity of a quantum sensor \cite{Poggiali-Cappellaro-Fabbri-2018PRX}, or a
combination of these requirements. Many useful tools in traditional
optimal control, such as the variational method, the Pontryagin
minimum principle and gradient algorithms, can be adapted to solving quantum optimal control problems. Optimal control techniques have been
widely applied to search for optimal pulses in physical chemistry
(for details, see, e.g., \cite{Rabitz-Vivie-Riedle-et-2000Sci}, \cite{Rice-and-Zhao-2000book}, \cite{Shapiro-and-Brumer-2003book}, \cite{Rabitz-Hsieh-and-Rosenthal-2004Sci}, \cite{Huang-and-Goan-2014PRA}, \cite{Shu-Ho-Xing-et-2016PRA}, \cite{Chakrabarti-and-Rabitz-2007IRPC}) and Nuclear Magnetic Resonance (NMR) experiments
\cite{Khaneja-Reiss-et-2003JMR}, \cite{Khaneja-Kehlet-et-2003PNAS},
\cite{Vandersypen-and-Chuang-2004RMP}. For example, GRadient Ascent Pulse Engineering (GRAPE) algorithms have been developed and applied in many NMR applications \cite{Khaneja-Reiss-et-2005JMR}. Time optimal control has been investigated for many spin systems theoretically as well as experimentally \cite{Wang-Allegra-et-2015PRL}, \cite{Geng-Wu-et-2016PRL}, which is important to achieve fast quantum gate operation in quantum computation to reduce the effect of quantum decoherence \cite{Cui-Xi-Pan-2008PRA}. An optimal control algorithm was designed based on the unconventional metric of sensitivity for quantum sensing and experimental results demonstrated that it improves the performance of a nitrogen-vacancy (NV) spin sensor \cite{Poggiali-Cappellaro-Fabbri-2018PRX}. Another example is quantum ensemble control. A quantum ensemble consists of a large number
of (e.g., up to $\sim 10^{23}$) single quantum systems (e.g., spin
systems) and in practical applications these individual systems in the ensemble could
have variations in the parameters that characterize the system
dynamics. It is generally impractical to employ different
control inputs for individual systems of a quantum ensemble. Hence, it is important to develop the means for
designing control fields that can simultaneously steer the ensemble
of systems from an initial state to a desired target state when
variations exist in the system parameters. A series of theoretical results on controllability of inhomogeneous quantum ensembles and systematic algorithms to achieve optimal control of inhomogeneous ensembles have been presented; e.g., see
\cite{Li-and-Khaneja-2006PRA}, \cite{Li-Ruths-et-2011PNAS}, \cite{Li-and-Khaneja-2009TAC}, \cite{Li-2011TAC}, \cite{Wang-and-Li-2018Auto}, \cite{Wang-and-Li-2017SIAM}, \cite{Turinici-2019PRA}, \cite{Chen-Dong-et-2014PRA}. This emerging topic originated from quantum control also inspires ensemble control with various interest and applications in both quantum and classical systems and control areas.

%[for details, see ,11,,108,109 of survey].
Another useful approach for constructing quantum open-loop controls is Lyapunov control.
The Lyapunov control methodology is a powerful tool for feedback
controller design in classical control systems. In quantum control,
it is usually difficult to obtain feedback information without affecting the system. In quantum Lyapunov control, we usually first complete the feedback control design by simulation to find a sequence of controls.
Then, we may apply the control sequence to the system under consideration in an open-loop form \cite{Mirrahimi-Rouchon-et-2005Auto},
\cite{Kuang-Dong-et-2017Auto}, \cite{Wang-and-Schirmer-2010A-TAC}. This provides a method of ``feedback design and open-loop control". The
most important tasks in quantum Lyapunov control involve the
selection of the Lyapunov function, the determination of the
control law and the analysis of asymptotic convergence.
For example, we may select the
following Lyapunov functions: $V_{1}(t)=\frac{1}{2}(1-|\langle
\psi_{f}|\psi(t)\rangle|^{2})$,
or $V_{2}(t)=\langle \psi(t)|Q|\psi(t)\rangle$, where $|\psi_{f}\rangle$ is
the target state and $Q$ is
a positive semidefinite Hermitian operator. The control function can be determined using the condition
$\dot{V}(t)\leq 0$ on the Lyapunov function $V(t)$ and LaSalle's invariance principle is useful
to analyze asymptotic convergence
\cite{D'Alessandro-2007Book}.
Lyapunov control techniques provide an intuitive strategy to design control laws for quantum systems with the possibility of guaranteeing asymptotic convergence for the control design algorithm. A limitation is that quantum Lyapunov control may converge slowly which is usually not expected due to fast time scale of quantum dynamics and unavoidable quantum decoherence. Rapid Lyapunov control and Lyapunov control with finite-time convergence are worth investigating following several preliminary results \cite{Kuang-Dong-et-2017Auto}, \cite{Hou-Khan-et-2012PRA}, \cite{Kuang-Guan-Dong-2021Auto}. Another potential limitation of quantum Lyapunov control is that the control law may not be robust against uncertainties and noise since the feedback design process usually assumes that the system model is exactly known.

\subsection{Quantum feedback control}
In classical control systems, a major aim of introducing feedback is to compensate for the effects of
unpredictable disturbances on a system under control, or to make
automatic control possible when the initial state of the system is
unknown. Feedback control has also been widely investigated for quantum systems. Measurement-based feedback control and coherent feedback control are two main classes of quantum feedback control strategies. In measurement-based feedback control, the information about the state of
system is obtained through measurement and then fed back for designing the control law. A challenge is that the process of obtaining feedback information may unavoidably disturb the state of the
system. Projective measurement and continuous weak
measurement are two main approaches to obtain feedback information
and the feedback controller usually consists of a classical system. The early development of quantum feedback control involved Markovian quantum feedback and Bayesian quantum feedback
\cite{Doherty-and-Jacobs-1999PRA}, \cite{Wiseman-and-Milburn-2010Book}.
In Markovian quantum feedback, the measurement record is immediately fed back onto the system
to alter the system dynamics and may then be forgotten \cite{Wiseman-Mancini-Wang-2002PRA} and the dynamics characterizing the resulting evolution
can be described by a Markovian master equation. Bayesian quantum feedback involves two steps of state estimation and
feedback control. The estimates of the dynamical variables are
obtained continuously from the measurement record, and fed back to
control the system dynamics \cite{Doherty-and-Jacobs-1999PRA}.
Bayesian feedback
usually has superior performance over Markovian feedback while it is more difficult to implement than Markovian feedback due to the existence of the
estimation step \cite{Wiseman-Mancini-Wang-2002PRA}.
Measurement-based quantum feedback can be used to improve the
control performance for e.g., quantum entanglement \cite{Yanagisawa-2006PRL}, \cite{Yamamoto-Nurdin-et-2008PRA}
quantum state preparation in various quantum systems such as quantum optical systems \cite{Wiseman-and-Milburn-2010Book}, \cite{Sayrin-Dotsenko-et-2011Nat}, and superconducting quantum
systems \cite{Vijay-Macklin-et-2012Nat}. The dynamics of quantum feedback control systems are often described by quantum stochastic master equations \cite{Qi-2009SciC} which can be obtained by quantum filtering theory. It is usually challenging to analyze and numerically simulate such stochastic master equations and characterize the effect of feedback time delay on control performance for complex quantum systems.

Quantum coherent feedback is a unique feedback control paradigm which has no counterpart for classical control systems. In quantum coherent feedback, the feedback controller itself is a quantum system which processes quantum information.
Quantum coherent feedback has been widely studied \cite{Lloyd-2000PRA}, \cite{Zhang-Wu-et-2012TAC}. For example, an $H^\infty$ control method has been used to investigate the robust control problem of linear quantum stochastic systems via quantum coherent feedback \cite{James-Nurdin-Petersen-2008TAC}. Mabuchi \cite{Mabuchi-2008PRA} demonstrated an experimental realization of a coherent feedback control system and verified the theory of linear quantum stochastic control in \cite{James-Nurdin-Petersen-2008TAC}. An LQG optimal control approach has been used to analyze quantum coherent feedback problems \cite{Nurdin-James-Petersen-2009Auto}.
These results mainly focus on linear quantum systems \cite{Nurdin-and-Yamamoto-2017Book}. Linear quantum systems are a class of quantum systems whose dynamics can be described by linear differential equations (usually in the Heisenberg picture) \cite{James-Nurdin-Petersen-2008TAC}, \cite{Petersen-2010MTNS}, \cite{Zhang-and-James-2011TAC}, \cite{Vladimirov-Petersen-James-2021AMO}. Many quantum optical systems, quantum superconducting circuits and optomechanical systems can be described or can be approximately characterized as linear quantum systems. Different from classical linear system equations, not every linear differential equation can correspond to a linear quantum system. Only such linear differential equations preserving specific commutation relations can be used to represent physical quantum systems or quantum controllers, which is referred as the problem of physical realizability in quantum control  \cite{James-Nurdin-Petersen-2008TAC}, \cite{Nurdin-and-Yamamoto-2017Book}. Various efforts have been made to develop linear quantum system theory; see, e.g., \cite{Petersen-2010MTNS}, \cite{Nurdin-and-Yamamoto-2017Book}, \cite{Zhang-and-James-2012CSB}, \cite{Zhang-Grivopoulos-Petersen-et-2018TAC}, \cite{Xiang-Petersen-Dong-2017Auto}. Another widely used modeling method for quantum coherent feedback is the $(S, L, H)$ method where $S$ is a scattering matrix, $L$ is a vector of
coupling operators and $H$ is a Hamiltonian operator \cite{Gough-and-James-2009TAC}, \cite{James-and-Gough-2010TAC}. An advantage of using a triple $(S, L, H)$ is that this framework automatically represents a physically realizable quantum system which has been widely used to describe open quantum systems and quantum network systems \cite{Gough-and-James-2009TAC}, \cite{Xiang-Petersen-Dong-2017TAC}. For example, robust stability and performance analysis of several classes of uncertain quantum systems have been investigated using the small gain theorem and the Popov approach based on the $(S, L, H)$ framework \cite{James-Petersen-Ugrinovskii-2013ACC}, \cite{Petersen-Ugrinovskii-James-2012PTRSA}, \cite{Xiang-Petersen-Dong-2017TAC}, \cite{Petersen-2017QST}.
 Quantum coherent feedback has also been used to control a single qubit in diamond \cite{Hirose-Cappellaro-2016Nat}, cool a quantum oscillator \cite{Hamerly-and-Mabuchi-2012PRL} and analyze quantum networks \cite{Zhang-and-Pan-2020Auto}. The performance of measurement-based feedback control and coherent feedback control has also been compared for different scenarios \cite{Jacobs-Wang-Wiseman-2014NJP}. A more comprehensive introduction to quantum feedback control can be found in \cite{Wiseman-and-Milburn-2010Book} and \cite{Zhang-Liu-et-2017PR}.

\section{Learning in quantum systems}
\subsection{Machine learning for quantum estimation and control}
Machine learning has been recognized as a powerful approach for solving many complicated quantum estimation and control problems. Learning control is one of the main control design approaches in quantum control and it has achieved great success in laser control of molecules and various other applications in quantum technology \cite{Judson-and-Rabitz-1992PRL}, \cite{Dong-2020Chapter}, \cite{Ma-and-Chen-2020SMC}, \cite{Niu-Boixo-et-2019NPJ}. Existing results of quantum learning control may be roughly classified into two classes: one class is quantum open-loop learning control and the other class is quantum closed-loop learning control. Quantum closed-loop learning control using genetic algorithms (GA) or differential evolution (DE) \cite{Dong-Xing-et-2020Cyber}
has achieved great successes in the laser
control of laboratory chemical reactions \cite{Rabitz-Vivie-Riedle-et-2000Sci},
\cite{Judson-and-Rabitz-1992PRL}. Closed-loop learning control
generally involves \cite{Rabitz-Vivie-Riedle-et-2000Sci}: (i) a trial laser control input design, (ii) the laboratory
generation of the control that is applied to the molecule samples, and (iii) a machine learning algorithm
that considers the prior experimental observation and suggests the
next control input. The initial trial input may be either a
well-designed control pulse or a random control input. The control objective is designed according to the practical requirement, which may be related to various factors such as the system's state, control energy and
control time, and is often mapped to the fitness function in the learning algorithm. In the learning process, a trial input is first applied to some samples and the result is observed. Second, a better control input is learned from a machine learning
algorithm based on the prior
experimental observation. Third, the ``better" control input is applied to new
samples. This process is repeatedly implemented until the control
objective is achieved or a termination condition is met. Closed-loop learning control is especially useful for quantum control problems where a large number of identical-state samples can be easily produced and the system dynamics are complicated. An example is to control chemical reactions of molecules using ultrafast laser pulses \cite{Rabitz-Vivie-Riedle-et-2000Sci} (e.g., fragmentation control of $\text{Pr(hfac)}_{3}$ \cite{Dong-Shu-et-2020TCST} and $\text{CH}_2\text{BrI}$ \cite{Dong-Xing-et-2020Cyber} molecules using femtosecond laser pulses). Closed-loop learning control provides a useful data-driven control method for guiding quantum physicists and quantum chemists to design efficient experiments, and there are many potential opportunities for systems \& control scientists and engineers to collaborate with them by developing efficient and robust quantum learning control algorithms.

In quantum open-loop learning control, quantum control tasks are formulated as an optimization problem and a machine learning algorithm may be employed to search for control inputs for solving the optimal control problem. In this method, we usually assume that there is known prior information about the system model and dynamics. Gradient algorithms are often incorporated into the development of quantum open-loop learning control due to their high efficiency \cite{Khaneja-Reiss-et-2005JMR}, \cite{Dong-Mabrok-et-2015TCST}, \cite{Dong-Shu-et-2020TCST}. In gradient algorithms, a critical task is to calculate the gradient information which may not be realistic to obtain for many practical applications. Moreover, some complex quantum control problems often have local optima. For these situations, stochastic search algorithms can be employed to search for good control fields with improved performance. For open-loop control, the learned control fields may not be robust enough for a quantum system with uncertainties and noise. Recent results show that it is possible to integrate some ideas from machine learning into the development of quantum control algorithms for achieving robust performance. For example, sampling-based learning control has been presented to achieve robust control design \cite{Dong-Mabrok-et-2015TCST} and several variants of GRAPE including a-GRAPE \cite{Ge-Ding-et-2020PRA}, b-GRAPE \cite{Wu-Ding-et-2019PRA} and d-GRAPE \cite{Wu-Chu-et-2018PRA} have been proposed to achieve high-precision and robust quantum control.

Machine learning can be used in quantum state estimation and parameter identification, and there are many opportunities to develop this area. For example, the powerful computational capability and generalizability of deep neural networks may provide an alternative solution to reconstructing quantum states and identifying parameters in quantum systems with noise. Machine learning can be integrated into adaptive strategies for achieving high-precision parameter identification. A challenge in quantum state estimation is that the complexity exponentially increases with the number of qubits, which imposes limitations on the applicability of machine learning to full reconstruction of quantum states. Another challenge is to characterize the optimal precision that can be achieved by a learning algorithm.

\subsection{Quantum machine learning}
Quantum machine learning explores how to design and implement quantum software to speed up machine learning and make it faster and more efficient than its counterpart on classical computers. Quantum machine learning usually makes use of quantum algorithms as part of its implementation and has the potential to outperform relevant classical machine learning algorithms for specific problems \cite{Biamonte-Wittek-et-2017Nat}. Various quantum machine learning algorithms such as quantum reinforcement learning, quantum principal component analysis, quantum support vector machines, and quantum neural networks have been presented (see, e.g., the survey papers \cite{Biamonte-Wittek-et-2017Nat}, \cite{Dunjko-and-Briegel-2018RPP} for more progress). Existing results show that there are quantum machine learning algorithms that exhibit quantum speedups. An example of quantum machine learning is quantum reinforcement learning where a quantum agent interacts with an environment and receives rewards for its actions \cite{Dong-Chen-et-2008TCyber-QRL}. The states and actions are represented by quantum superposition states and the reinforcement strategy can be designed based on quantum measurement and Grover operation. It has been theoretically proven that quantum reinforcement learning may achieve quadratic speedup over its classical counterpart \cite{Dunjko-Taylor-Briegel-2016PRL}, \cite{Paparo-Dunjko-et-2014PRX}. A quantum speedup of the learning time for an agent by a quantum communication channel with the environment has been experimentally demonstrated \cite{Saggio-Asenbeck-Hamann-et-2021-Nat}, and quantum reinforcement learning has also been used to explain human decision-making process \cite{Li-Dong-et-2020NatHB}. A quantum algorithm for solving a linear set of equations over $2^{n}$-dimensional vector spaces in time that is polynomial in $n$ has been presented (also called the HHL algorithm), which implements exponentially faster than its best known classical counterparts \cite{Harrow-Hassidim-Lloyd-2009PRL}, \cite{Biamonte-Wittek-et-2017Nat}. The key idea behind this algorithm is that the quantum state of $n$ qubits is a vector in a $2^{n}$-dimensional complex vector space and quantum mechanics is all about matrix operations on vectors in high-dimensional vector spaces \cite{Biamonte-Wittek-et-2017Nat}. HHL has been employed as a subroutine to develop various quantum machine learning algorithms such as Bayesian inference, quantum principal component analysis \cite{Lloyd-Mohseni-et-2014NPhy} and quantum support vector machines. Another promising quantum machine learning approach involves quantum neural networks which aim to develop quantum counterparts of neural networks and deep learning. Quantum information processing provides new models and features (e.g., quantum entanglement, quantum coherence) for developing novel quantum deep learning which may be more efficient than classical deep learning. For example, quantum coherence can quadratically reduce the number of samples needed to learn a desired task when training a quantum Boltzmann machine \cite{Biamonte-Wittek-et-2017Nat}. As another recent progress, variational quantum algorithms are widely investigated for achieving quantum advantage using near-term quantum computers where a task is usually encoded into a parameterized cost function which is evaluated using a quantum computer, and the parameters are trained by a classical optimizer \cite{Cerezo-Arrasmith-Babbush-et-2021NRP}. The framework of variational quantum algorithms has also been employed for developing quantum machine learning applications. The development of quantum machine learning is still at its early stage and there are many opportunities to make contributions to this emerging area. Several challenging examples include how to characterize the cost of inputs and the complexity of quantum gates for a quantum machine learning algorithm, how to benchmark the performance of a quantum machine learning algorithm against its classical counterpart, and how to implement various quantum machine learning algorithms on an experimental quantum system. Besides these opportunities of further developing quantum machine learning algorithms and quantum learning theory, there are plenty of opportunities to explore the possibility and potential of using quantum machine learning to efficiently solve some challenging estimation and control problems in traditional systems and control community, and to develop quantum machine learning solutions to challenging quantum state estimation, parameter identification and control design problems.

\section{Concluding remarks}
Estimation, control and learning of quantum systems have become critical drives to facilitate the emerging area of quantum technology, which present many new challenging problems different from classical systems and also provide a lot of exciting opportunities for researchers in the area of systems and control. Although some results have been presented, many topics of quantum systems and control are worth further exploring. For example, robustness and reliability are two key requirements for practical quantum technology, and quantum robust control methods such as $H^\infty$ control \cite{James-Nurdin-Petersen-2008TAC}, noise filtering \cite{Soare-Ball-et-2014NP}, sliding mode control \cite{Dong-and-Petersen-2012Auto}, and fault-tolerant control \cite{Wang-and-Dong-2017TAC}, \cite{Liu-Dong-et-2021TAC} should be further developed. Adaptive and optimal control strategies could be further investigated to enhance the sensitivity of emerging quantum sensors \cite{Bao-Qi-Dong-2022PRApplied} and the precision of quantum parameter identification and quantum detector tomography \cite{Wang-Yokoyama-Dong-et-2021TIT}. The applicability of quantum observability theory \cite{Miao-James-Petersen-2016Auto} and its difference from classical observability theory could be further explored. Efficient estimation and control algorithms need to be further developed for high-dimensional quantum systems, quantum networks and attosecond dynamical systems \cite{Shu-Guo-Yuan-et-2020OL}. It needs the collaborative efforts of researchers from various areas of systems and control, quantum physics, quantum engineering, physical chemistry and quantum information to systematically solve these challenging problems and accelerate the development of quantum technology.

\section*{References}
%
%\end{thebibliography}
\bibliographystyle{alpha}

\end{document}